\documentclass[11pt,a4paper]{article}
\pdfoutput=1
\usepackage{jheppub}
\usepackage{dsfont}

\usepackage{mqcommands}

\newcommand{\gn}[1]{\Gamma\left(#1\right)}
\newcommand{\gt}[1]{\Gamma\left(\frac{#1}2\right)}
\newcommand{\reef}[1]{(\ref{#1})}

\newcommand{\iads}{\int_{\mbox{\tiny AdS}}}

\title{Loops, Polytopes and Splines}

\author{Miguel F. Paulos}
\affiliation{Department of Physics, Brown University, Box 1843, Providence, RI 02912-1843, USA}

\emailAdd{miguel\_paulos@brown.edu}

\abstract{
We uncover an unexpected connection between the physics of loop integrals and the mathematics of spline functions. One loop integrands are Laplace transforms of splines. This clarifies the geometry of the associated loop integrals, since a $n$-node spline has support on a $n$-vertex polyhedral cone. One-loop integrals are integrals of splines on a hyperbolic slice of the cone, yielding polytopes in $AdS$ space. Splines thus give a geometrical counterpart to the rational function identities at the level of the integrand. Spline technology also allows for a clear, simple, algebraic decomposition of higher point loop integrals in lower dimensional kinematics in terms of lower point integrals - e.g. an hexagon integral in 2d kinematics can be written as a sum of scalar boxes. Higher loops can also be understood directly in terms of splines - they map onto spline convolutions, leading to an  intriguing representation in terms of hyperbolic simplices integrated over other hyperbolic simplices.  We finish with speculations on the interpretation of one-loop integrals as partition functions, inspired by the use of splines in counting points in polytopes.
}

\begin{document}
\maketitle 

\section{Introduction}

Scattering amplitudes are beautiful multi-faceted objects. Nowhere does this show more clearly than in the maximally supersymmetric $SU(N)$ Yang-Mills theory ~\cite{Brink:1976bc}, especially in the planar ($N\to \infty$) sector \cite{SCATReview,INTReview}. There, new perspectives on scattering amplitudes as expectation values of null-polygon shaped Wilson loops \cite{Alday:2007hr} eventually led to the discovery of dual conformal symmetry \cite{Drummond:2007cf,Drummond:2007au,Drummond:2008vq} and the consequent infinite symmetry algebra known as the Yangian \cite{Drummond:2009fd}. Relatedly, scattering amplitudes have been shown to be intimately connected to correlation functions of stress-tensor multiplets \cite{Eden:2011yp,Eden:2011ku}. Concurrently, another intriguing picture seems to be emerging connecting the Yangian invariants to the algebra and combinatorics of the positive part of the Grassmannian - the space of $k\times n$-dimensional planes which satisfy certain positivity conditions \cite{ArkaniHamed:2009dg,ArkaniHamed:2009sx,ArkaniHamed:2009vw,ArkaniHamed:2009dn,Postnikov}.

The wealth of riches does not stop here. Mellin space \cite{Symanzik:1972wj,Mack:2009gy,Mack:2009mi}
is the natural place to discuss conformal correlation functions, and has led not only to major technical progress but also to new perspectives on bulk locality and the flat-space limit of $AdS$ scattering amplitudes \cite{Penedones:2010ue,Fitzpatrick:2011ia,Paulos:2011ie,Nandan:2011wc,Fitzpatrick:2011dm,Fitzpatrick:2011hu,Fitzpatrick:2012cg}. Whether correlation functions are computed from an $AdS$ dual \cite{Witten:1998qj,Aharony:1999ti} or directly in flat-space \cite{Paulos:2012nu}, a perturbative formulation ensures that their Mellin amplitudes take a simple form highly reminiscent of scattering amplitudes themselves. Mellin space also sheds light on the various identities that have been uncovered relating different kinds of conformal loop integrals. For instance, in \cite{Paulos:2012nu} a simple formula was written down connecting the scalar double box integral to the $d=6$ hexagon. Many other such identities can be written down, owing to the particular fact that Mellin amplitudes of flat space correlation functions seem to satisfy Feynman rules. In particular, Mellin amplitudes for position space tree-level diagrams take a factorized form, which implies that the corresponding position space formulae take the form of convolutions of contact interactions. The above mentioned identity is the simplest case for this, with a $3\to 3$ exchange diagram being written in terms of a $6$-pt contact interaction\footnote{In the dual momentum space, these are scalar double box and hexagon integrals, respectively}. Analogously, one could write a ladder diagram with $n$-boxes in terms of a simple $(n-1)$-fold integral of a $4+2n$ contact interaction. It is therefore especially important to understand the latter integrals. They can be thought of as $4+2n$-gon integrals in as many dimensions, dimensionally reduced to $4$ dimensions, and a systematic understanding of these was one of the main motivations for this paper.

Curiously, the answer to this problem arose from an examination of yet another perspective on loop integrals, which connects scattering amplitudes to {\em convex geometry}. This has so far been seen to some extent from two different sides, both at level of the {\em integrand} \cite{Hodges:2009hk,ArkaniHamed:2010gg,ArkaniHamed:2010gh,ArkaniHamed:2010kv}, as well as the {\em integral} \cite{Mason:2010pg}. In the first, the BCFW \cite{Britto:2005fq} recursion relation acquires a remarkable interpretation, that of a simplicial decomposition of volumes of polytopes in projective space. In the second, the integrals were seen to compute volumes of polytopes, or boundaries thereof, in hyperbolic space.

It is with this last approach that this note is concerned with. We examine loop integrals, conformal and otherwise, from a new perspective, connecting loop integrals to the mathematics of splines\footnote{The literature on splines is vast and we cannot do justice to it here. We have found the book \cite{hyperplane} and articles \cite{Carlson, Concini} especially useful and a large list of references can be found there.} - a special class of functions better known for its uses in {\em approximation} theory! The splines have positivity and convexity built in and provide the missing link in connecting the integrand and the integral. There are several kinds of splines. The ones we will concern ourselves with are sometimes called multivariate truncated powers, and they are piecewise polynomials with support on polyhedral cones, whose vertices correspond to vectors in projective space. The loop integral integrates over the spline with a gaussian measure, reducing the integral from the cone to a hyperbolic slice thereof, leading to the appearance of polyhedra in $AdS$. It is important to emphasize that our methods are in no way restricted to $N=4$ SYM - they are universal and apply to any quantum field theory, conformal or otherwise, and are really statements on integrals of products of propagators. In particular, this holds even for the calculation of Witten diagrams in $AdS$ space! In this way we will find a beautiful generalization of the star-triangle relation, where many-legged $AdS$ stars get glued up into $AdS$ polyhedra. Crucial to our approach will be the use of the embedding space formalism: the spline naturally lives in this space and it is from there that the polyhedra descend.

Let us discuss the outline of this paper. After some preliminaries in section \ref{ambientspace} we move on to section \ref{warmup}, where we give a new point of view on the old familiar scalar box integral. This allows us to introduce the spline in a simple case which captures most of the important geometric features of more complicated integrals. Section \ref{generalizedstartriangle} greatly generalizes these results, showing that hyperbolic polyhedra hide in disguise in a much broader class of integrals, among which we include contact interactions in $AdS$. Accordingly we get a nice map from $AdS$ stars into $AdS$ polyhedra. In section \ref{laplace} we emphasize that the spline is nothing but the Laplace transform of the one-loop integrand. This allows one to trade identities of rational functions for geometric ones, and vice versa. We examine in particular how the boundary of a four-simplex can be alternatively be given as a pentagon integral or as a sum of boxes, clarifying the simple representations found in \cite{ArkaniHamed:2010gh}. The following section deals with the calculation of splines which have large numbers of nodes. What this means is that geometrically these correspond to splines associated with non-simplicial cones; from the point of view of the integrand it corresponds to considering loop integrals involving a number of external points greater than $d+2$. We present two methods for computing the spline in these cases, one intuitively obvious in terms of partial fraction decomposition of the integrand; and another, more powerful in terms of a residue computation. As a simple application, we show how to predict the result for a box integral in 1d kinematics from a purely geometric perspective. We comment on more advanced applications, but leave detailed calculations for a future publication \cite{workinprogress}. Next we move on to considerations of higher loop integrals. We begin by showing that one-loop integrals can be written in several different equivalent ways, and how one spline can actually be decomposed into two or more. Then we perform an explicit two-loop computation which shows a very similar structure. In both cases we show that the spline picture takes the form of integrals of hyperbolic simplices over other simplices. This shows that the spline picture is not intrinsic to one-loop computations, although the physical meaning of this picture is far from obvious. We finish with a discussion and some speculations, in particular on the idea of the spline as a charge degeneracy. The paper is complemented by two appendices with some technical details on the two-loop calculation and the link between the spline we use and Carlson's Dirichlet B-splines \cite{Carlson}.

\section{Set-up}
\label{ambientspace}

\subsection{Ambient space}

In the calculations to follow, it is crucial that they should be seen in the light of the embedding or ambient space formalism ~\cite{Dirac:1936fq,Weinberg:2010fx,Costa:2011dw,Costa:2011mg}. In this approach we think of $d$-dimensional Euclidean space as a slice of the projective light-cone in $D=d+2$ dimensions. The $SO(d+1,1)$ invariance group acts linearly on $D$ dimensional vectors and induces a non-linear representation of the conformal group in $d$ dimensions. We will consider null vectors in $D$ dimensions which we will invariably denote by capital letters $P$ and $Q$, the latter being used for dummy variables. These are projective null vectors which means we have
	\be
	P^M P_M=-P^+ P^-+P^\mu P_\mu=0, \qquad P\simeq \lambda P
	\ee
where $M, \mu$ are $D$ and $d$-dimensional indices respectively.

We define a reference vector $I$, or infinity vector, which explicitly breaks conformal $SO(d+1,1)$ invariance. The vector $P^M/(-P\cdot I)$ parameterizes $d$-dimensional flat space. We choose the gauge $-P\cdot I=\frac{\sqrt{2}}2 P^+$, such that
	\be
	\frac{P^M}{-P\cdot I}=\sqrt{2}\, (1,x^2, x^\mu),
	\ee
By contracting two independent $P$ vectors we obtain
	\be
	P_{ij}\equiv \frac{(-P_i \cdot P_j)}{(-P_i \cdot I)(-P_j \cdot I)}=(x_i-x_j)^2.
	\ee
The factors $P_i\cdot I$ drop out of conformally invariant expressions, and we can in any case recover them by demanding homogeneity zero in rescalings of the $P_i$.

So much for null vectors. We will also need to consider massive vectors, satisfying $X^2=-1$. These parameterize a $d+1$-dimensional $AdS$ space and can be written
	\bea
X=\frac{1}{x_0}\left(1,x_0^2+x^2,x^\mu\right).
	\eea
The coordinates $x_0$ and $x^\mu$ are simply those in the usual parameterization of the Poincar\'e patch of $AdS$,
	\bea
	ds^2_{\mbox{\tiny AdS}}=\frac{\ud x_0^2+\ud x\cdot \ud x}{x_0^2}.
	\eea
\subsection{Dual conformal symmetry}

\begin{figure}
	\centering
		\includegraphics[scale=1.5]{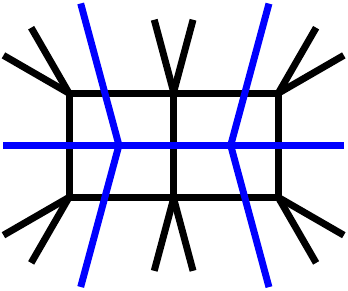}
	\caption{A fully massive two-loop scalar box integral (black) and its position space counterpart (blue).}
	\label{fig:2loop}
\end{figure}

As mentioned in the introduction, $N=4$ SYM has a dual conformal symmetry. We will not focus particularly on $N=4$ in the following, but we will be using the idea that the same integral can be thought of in two dual manners. What this means in practice is that if one does the replacements
	\bea
	p_i\to x_i-x_j
	\eea
in a momentum space planar loop integral, the result looks like the calculation of a position space correlation function. This is illustrated for the double box integral in figure \ref{fig:2loop}.

In practice we are replacing propagators as
	\bea
	\frac{1}{p_i^2}\rightarrow \frac{1}{(x_i-x_{i+1})^2}=\frac{1}{P_{ij}}
	\eea
It is from this position space perspective that we will be working. We are {\em not} following the convention that $(x_i-x_{i+1})^2=0$. Some of the integrals we will consider will have no interpretation from the momentum space point of view, since we will consider propagator factors such as $1/(x_1-x_2)^{2\Delta}$ with $\Delta\neq 1$. These do appear however in $AdS$/CFT calculations. We will keep an attentive eye regarding applications to momentum space loop integrals. 

We shall also consider massive momentum space loop integrals. To what does this correspond in the dual space? Well, such a propagator is given by
	\bea
	\frac{1}{p^2+m^2}\to \frac{1}{(x-y)^2+m^2}=\frac{1}{m}\, \frac{m}{(x-y)^2+m^2}
	\eea
This suggests we identify $m$ as the $x_0$ coordinate of an $AdS$ vector $X$, since we have e.g.
	\bea
	\frac{1}{P_i\cdot X_j}=\frac{x_0}{(x_i-x_j)^2+x_0^2}.
	\eea
The mass then becomes the coordinate $x_0$ in $AdS$. In this note we will happily consider massless vectors $P$ or massive vectors $X$ as our external data. When a dual momentum space perspective is warranted, these correspond to loop integrals with massive or massless propagators.  In practice, throughout this paper we will be thinking of the integrals as position space tree-level correlators. Nevertheless, we will keep the convention of denoting the number of integrals involved by the loop-order - which equals the number of interaction vertices in the position space picture. 

\section{Warm-up: the box and the spline}
\label{warmup}
As a warm-up to greater things to come, let us consider the familiar scalar box integral. In the dual conformal picture, this is the same as a position space contact interaction. In any case, in the embedding space formalism we can write this integral as
	\bea
	{\hbox{\lower 12pt\hbox{
\begin{picture}(30,30)
\put(0,0){\includegraphics[width=30pt]{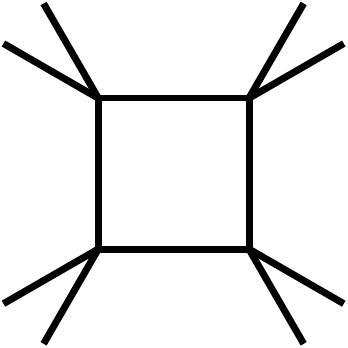}}
\end{picture}
}}}=\int \frac{\ud^4 Q}{2\pi^2}\, \frac{1}{(P_1\cdot Q)\,(P_2\cdot Q)\, (P_3\cdot Q)\, (P_4 \cdot Q)},
	\eea
with $\ud^4 Q\equiv \ud^4 x$. We begin by following the standard route for evaluating such integrals: we introduce Schwinger parameters to exponentiate the propagators, obtaining
	\bea
\int \frac{\ud^4 Q}{2\pi^2}\, \int_0^{+\infty} \prod_{i=1}^4 \ud t_i\, e^{(\sum_{i=1}^4 t_i P_i)\cdot Q}.
	\eea
After performing  the $Q$ integration (which is simply a gaussian integral), we get
	\bea
	\int_0^{+\infty} \prod_{i=1}^4 \ud t_i\, e^{(\sum_{i=1}^4 t_i P_i)^2}.
	\eea
Up to now our manipulations have been standard, but now we do a trick, writing
	\bea
	{\hbox{\lower 12pt\hbox{
\begin{picture}(30,30)
\put(0,0){\includegraphics[width=30pt]{1loopmassive.pdf}}
\end{picture}
}}}=\int_{\mathds M^{6}} \!\!\ud X\, e^{X^2}\, \mathcal T(X;P_i)
	\eea
where $\mathds M^6$ is 6-dimensional Minkowski space, and the function $\mathcal T(X;\{P_i\})$ is defined by:
	\bea
	\mathcal T(X;\{P_i\})=\int_0^{+\infty} \prod_{i=1}^4 \ud t_i \, \delta(X-\sum_{i=1}^4\, t_i P_i). \label{splinebox}
	\eea
The function $\mathcal T$ is known as the {\em multivariate spline}, or truncated power function. Here we shall simply call it the spline. 
Splines are best known for their applications in approximation theory, computer graphics, and also for the important problem of counting integer points in polytopes - we will comment on this latter application in the discussion. Here we see them making an unexpected appearance in quantum field theory! A spline is a function depending on a set of vectors, here the $P_i$, which are known as its nodes. There can be any number of these, but if the number of nodes is less than the dimensionality of $X$ (in this case, six), the spline is not a function but a tempered distribution. To compute the spline, let us assume that the vectors $P_i$ are generic, i.e. there are no linear degeneracies between them. Now introduce two other vectors $P_5, P_6$ such that the full set of the $P_i$ forms a basis of $\mathds M^{6}$. We can then write the spline as
	\bea
	\mathcal T(X;\{P_i\})=\int_0^{+\infty} \prod_{i=1}^6 \ud t_i\, \delta(t_5)\,\delta(t_6)\, \delta(X-\sum_{i=1}^6 t_i P_i).
	\eea
By thinking of the set of six $P_i$ vectors as forming the columns of a matrix $M$, the set of $t_i$ integrals is easily performed: the six dimensional delta function imposes $t_i=W_i\cdot X$, with
	\bea
	W_i^M=\frac{1}{5!\sqrt{\det P_{ij}}}\epsilon^M_{\  N_1\ldots N_{5}}\epsilon_i^{\ j_1\ldots j_{5}}\,
	P_{j_1}^{N_1}\ldots P_{j_{5}}^{N_{5}}.
	\eea
Notice that the $W_i^M$ are simply the rows of $M^{-1}$. There is one catch though: since the $t$ variables are positive, this imposes constraints on $W_i\cdot X$. The result is then
	\bea
	\mathcal T(X;P_i)=\frac{\delta(W_5\cdot X)\delta(W_6\cdot X) \prod_{i=1}^4 \Theta(W_i\cdot X)\,}{\sqrt{\det{P_{ij}}}},\label{splinecomputed}
	\eea
with $\Theta(x)$ the Heaviside Theta function, and $\sqrt{\det{P_{ij}}}=\det M$.

This result for the spline has a very simple and beautiful geometric interpretation. Going back to equation \reef{splinebox} we see that the spline can only be non-zero if $X$ can be written as a positive linear combination of the vectors $P_i$; such positive linear combinations form a pointed\footnote{It is pointed because for all $P_i$ we have $P_i^+=\sqrt{2}$, hence the origin does not belong to the cone.} polyhedral cone, and hence the spline has support only in the interior of this cone. The numerator in equation
\reef{splinecomputed} is precisely the characteristic function\footnote{The characteristic function $\chi_S(x)$ of a set $S$ is defined as being unity if $x\in S$, zero otherwise.} for this polyhedral cone. Indeed, the $W_i$ vectors define hyperplanes on which all vectors except $P_i$ lie - forming the faces of the polyhedral cone.

\begin{figure}
	\centering
	\includegraphics[scale=0.5]{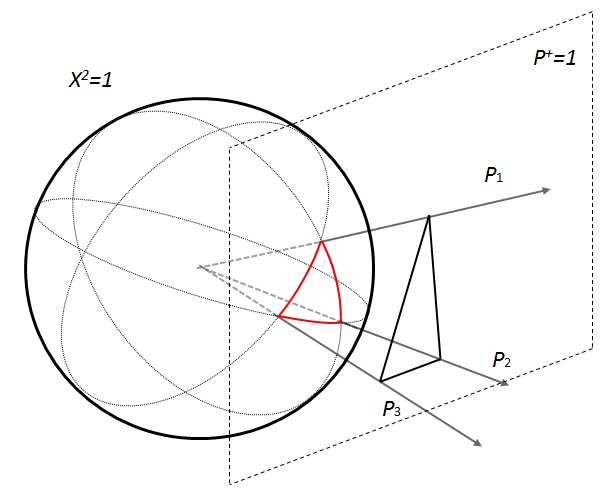}
	\caption{A projective triangle and its spherical shadow. A similar picture holds in the hyperbolic case and with tetrahedra (though harder to visualize).The projective triangle lies on the $P^+=1$ hyperplane.}
	\label{shadow}
\end{figure}

Let us go back to the computation of the box integral. With our result for $\mathcal T$ we see that it reduces to a gaussian integral over the polyhedral cone. But we can do better! Since the gaussian measure only cares about the norm of $X$, we can do this integral straightaway. Indeed, since the spline is homogeneous of degree $-2$ in $|X|$ this integral can be done quite easily, yielding a gamma function. Actually, there is an important point: since the six-dimensional space has necessarily Lorentzian signature, the sign of $X^2$ isn't necessarily fixed. However, since all the $P_i$ are future pointing, any positive linear combination of them will form a timelike vector, and hence $X^2<0$ inside the cone. After the integration over the scale of $X$ is made we are therefore left with an integral over vectors inside the polyhedral cone which further satisfy $X^2=-1$ - in other words, over an $AdS$ slice of the polyhedral cone. In this case, such a slice is a hyperbolic ideal\footnote{Here ideal refers to the fact that its vertices lie on the $AdS$ boundary.} tetrahedron. Let us denote such tetrahedron by $\mathcal P_4$. Then we have just shown that
	\bea
	{\hbox{\lower 12pt\hbox{
\begin{picture}(30,30)
\put(0,0){\includegraphics[width=30pt]{1loopmassive.pdf}}
\end{picture}
}}}=\frac{1}2\,\iads\!\! \ud X\, \mathcal T(X;P_i)=\frac{1}2\,  \frac{\mbox{Vol($\mathcal P_4$)}}{\sqrt{\det P_{ij}}}
	\eea
The $X$ integration measure is such that the denominator above only involves the $P_i$ with $i=1,\ldots 4$. The box integral is equal to a volume ratio; in the denominator we have the volume in $\mathds CP^5$ of the tetrahedron with vertices denoted by the $P_i$, volume which is computed with respect to the reference vector $I$. In the numerator we have the volume of the hyperbolic ``shadow'' cast by that solid: the hyperbolic 3-simplex $\mathcal P_4$. This is illustrated in figure \ref{shadow}. 

For a one-loop integral such as the one we have just computed, it is quite natural to change the normalization so that the denominator factor changes. However we shall see later on that in more general computations the natural objects to consider are precisely these volume ratios; or put differently, $AdS$ integrals of splines.

\section{Generalized Star-Triangle relations}
\label{generalizedstartriangle}
While this interpretation of the box integral as a volume in $AdS$ is not new, and was in fact derived fairly recently in \cite{Mason:2010pg}, our way of obtaining it certainly is, and sheds light on the mechanism at work: the essential point is that one-loop {\em integrands} are Laplace transforms of splines, as we shall see in section \ref{laplace}. This fact immediately allows for many generalizations to other kinds of loop integrals.

Let us therefore consider a more general class of one-loop integrals\footnote{This class of integrals was first studied in \cite{Symanzik:1972wj} for the special case $m_i=0$ and $\sum \Delta_i=d$. See also \cite{Paulos:2012nu} for more recent work.} , the colored polygons in arbitrary dimension:
	\bea
	I=\int \frac{\ud^d Q}{2\pi^{d/2}}\, \prod_{i=1}^n \Gamma(\Delta_i)\, \frac{m_i^{\Delta_i}}{(P_i\cdot Q_i)^{\Delta_i}} \label{polygons}
	\eea
The colouring refers to the fact that each vertex $P_i$ now has an associated weight, $\Delta_i$, or ``colouring''. We also relax the constraint that $P_i$ is null, allowing for $P_i^2=-m_i^2$. Accordingly this implies that $P_i/m_i\equiv X_i$ is a vector lying on an $AdS$ subspace of $\mathds M^{d+2}$, with $m\equiv x_0$. Accordingly, in \reef{polygons}, the normalization has been chosen such that we can trade $P/m$ by $X$. For all $\Delta_i=1$ the above integral can be understood in the dual conformal picture as a one-loop massive polygon integral. Non-unity values of $\Delta_i$ arise frequently in conformal integrals. Here we show that this entire class is associated with colored simplices or polytopes in $AdS_{d+1}$.

The computation is straightforward, and is entirely analogous to the one in the previous section. We introduce Schwinger parameters and perform the $Q$ integration which is simply a gaussian. Generically the conformality condition $\Sigma\equiv \sum_{i=1}^n \Delta_i=d$ can be violated. The result is
	\bea
	I=\int_0^{+\infty} D^n_\Delta t\, (\sum t_i)^{\Sigma-d}\, e^{(\sum t_i X_i)^2}
	=\int_{\mathds M^D} \ud X (-\sqrt{2}I\cdot X)^{\Sigma-d}\, e^{X^2}\, \mathcal T_{\{\Delta_i\}}(X;\{X_i\})
	\eea
where we have introduced for convenience the measure $D^n_\Delta t\equiv \prod_{i=1}^n \ud t_i\, t_i^{\Delta_i-1} $, and the generalized version of the spline is
	\bea
	\mathcal T_{\{\Delta_i\}}(X;\{X_i\})=\int_0^{+\infty} D^n_\Delta t\,\,\delta(X-\sum_{i=1}^n t_i X_i). \label{genspline}
	\eea
Notice that the spline now depends on an extra set of arguments, namely the weights $\Delta_i$ associated with each node. We will omit this argument in the case where all $\Delta_i=1$.
As before, the spline has support inside the polyhedral cone with vertices $X_i$, and is homogeneous in $|X|$ with degree $\Sigma-D$.
The integral over the norm $|X|$ is again performed trivially and one is left with
	\bea
	I=\frac{\Gamma\left(\Sigma-h\right)}{2}\,\iads \ud X\, (-\sqrt{2}\,I\cdot X)^{\Sigma-d}\, \mathcal T_{\{\Delta_i\}}(X;\{X_i\})
	\eea
After the integration over the norm the integral is now over a colored polyhedron whose vertices lie on the horospheres with radii $m_i$ in $AdS$, where again the coloring refers to the fact that the vertices of this polyhedron are labelled by the weights $\Delta_i$. But what are we integrating over? To see this we need to evaluate the spline explicitly and this is done starting from expression \reef{genspline}. The final result depends crucially on the number $n$ of nodes and on the weights $\Delta_i$. We will address the cases $n>D$ in section \ref{simplicial}. The simplest case is when $n=D$, whereupon the spline becomes
	\bea
	\mathcal T_{\{\Delta_i\}}(X;\{X_i\})=\frac{\prod_{i=1}^D (W_i\cdot X)^{\Delta_i-1}\,\Theta(W_i\cdot X)}{\sqrt{\det X_i\cdot X_j}}
	\eea
Just like in the previous section, the $\Theta$ functions enforce that the spline is supported only inside the hyperbolic polyhedron $\mathcal P$ with vertices $X_i$, or equivalently in the interior of the hyperplanes defined by the $W_i$. The case where the number of nodes is less than $D=d+2$ is also simple: we just eliminate the corresponding $W_i\cdot X$ factor and replace the Heaviside Theta function by a Dirac delta function, which further localizes the spline. In any event, both these cases lead to
	\bea
	I=\frac{\Gamma\left(\Sigma-h\right)}{2\sqrt{\det X_i\cdot X_j}}\int_{\mathcal P_n} \ud X\, (-\sqrt{2}\,I\cdot X)^{\Sigma-d}\, \prod_{i=1}^n (W_i\cdot X)^{\Delta_i-1}.
	\eea
That is, the integral is interpreted as the computation of certain ``moments'' of a hyperbolic polyhedron $\mathcal P_n$. A direct interpretation of the integral as a volume is only possible in the special case $\Sigma=d$ and all $\Delta_i=1$ - nevertheless there is still an underlying geometrical structure. In any case, $I$ can still be considered as an integral of certain homogeneous factors over a hyperbolic polyhedron. Using the alternative form
	\bea
	I=\int \ud X e^{X^2} (-\sqrt{2}\,I\cdot X)^{\Sigma-d} \frac{\prod_{i=1}^D (W_i\cdot X)^{\Delta_i-1}\,\Theta(W_i\cdot X)}{\sqrt{\det X_i\cdot X_j}},
	\eea
we see that something quite interesting occurs if all dimensions $\Delta_i$ are integers. Indeed, in this case the presence of the gaussian measure allows us to integrate by parts. This gets rid of the annoying $X$ factors and at the same time the derivatives hit the $\Theta$ functions, further collapsing the integral onto lower point simplices. This allows for a geometrical interpretation of these integrals as sums of volumes with well determined coefficients. We will see some concrete examples of this in section \ref{simplicial}. 

One interesting outcome of these results, is that we see that even non-conformal integrals (with arbitrary $\Delta_i$) can suitably interpreted as integrals in $AdS$. But there is a different way of seeing it. With the parameterization,
	\bea
	X=\frac{1}{x_0}(1,x_0^2+x^2,x^\mu)
	\eea
we have $X\cdot I=1/x_0$ and the $AdS$ measure is $\ud X=\ud x^\mu\, \frac{\ud x_0}{x_0^{d+1}}$. Therefore the factors $(X\cdot I)^{\Sigma-d}$ can perhaps be better thought of as deforming the $AdS$ measure into $\ud x^\mu\, \frac{\ud x_0}{x_0^{\Sigma+1}}$, thereby breaking conformality.

Let us mention another possible generalization of our results. The appearance of the spline can be seen to be entirely due to the factorized form of the integrand. Such factorized forms appear not only in flat-space integrals, but also in the computation of $AdS$ Witten diagrams \cite{Witten:1998qj}, and in particular of contact interactions, which take the form:
	\bea
	\int \ud X\, \prod_{i=1}^n \frac{\Gamma(\Delta_i)}{(P_i\cdot X)^{\Delta_i}}
	\eea
The only difference from the flat-space case is that now the result is automatically conformal due to the different integration measure, and hence the awkward factors of $I\cdot X$ never appear. We come then to the beautiful result that there is a generalized ``star-triangle'' relation in $AdS$ space: star-like contact interactions get glued up into integrals over polyhedra in $AdS$, exchanging vertices and hyperplanes:
	\bea
	\iads \!\!\ud X\, \prod_{i=1}^n \frac{\Gamma(\Delta_i)}{(P_i\cdot X)^\Delta_i}\quad \simeq \quad \iads\!\! \ud X \prod_{i=1}^n\frac{\Theta(W_i\cdot X)}{(W_i\cdot X)^{1-\Delta_i}}
	\eea
In section \ref{multiloop} we will see that this theme persists even beyond the one-loop/contact interaction level.

\section{From rational functions to geometry and back again}
\label{laplace}

In the previous sections we have seen how the factorized form of the integrand is directly connected to its representation in terms of the spline. In fact, we can make a stronger statement than this, namely that the one-loop integrand is the Laplace transform of the spline. Indeed, we have
	\bea
	\prod_{i=1}^n \frac{\Gamma(\Delta_i)}{(P_i\cdot Q)^{\Delta_i}}=\int_0^{+\infty}D^n_{\Delta} t\, e^{Q\cdot (\sum_i t_i P_i)}=\int_{\mathds M^D} \ud X\, e^{Q\cdot X}\, \mathcal T_{\{\Delta_i\}}(X;\{P_i\}).
	\eea
This simple fact has important consequences, since it allows us to translate identities between rational functions, at the level of the integrand, into geometrical identities in the spline representation.

Let us start with a trivial example. For concreteness we fix $d=n=4$ and set all $\Delta_i=1$. Then we have the trivial identity
	\bea
	\frac{1}{\prod_{i=1}^4 P_i\cdot Q}=\frac{P_5\cdot Q}{\prod_{i=1}^5 P_i\cdot Q}
	\eea
Taking the Laplace transform this translates into
	\bea
	\frac{P_5\cdot Q}{\prod_{i=1}^5 P_i\cdot Q}&=&\int_{\mathds R^D} \ud X e^{Q\cdot X}\, P_5\cdot(-\partial_X)\, \mathcal T_{\{\Delta\}}(X,\{P_i\})\nonumber \\
	&=&-
	\int_{\mathds R^D} \ud X e^{Q\cdot X}\, (P_5\cdot \partial_X) \left(\frac{\prod_{i=1}^5 \Theta(W_i\cdot X) \delta(W_6\cdot X)}{\sqrt{\det P_i\cdot P_j}}\right)
	\eea
where again we have introduced an extra vector $P_6$. The intepretation of this equation is straightforward: four nodes correspond to a tetrahedron; but a tetrahedron can also be thought of as one of the faces of a 4-simplex, which is part of its boundary. It is this fact that is translated mathematically by the right-hand-side of the above equation. Indeed, the derivative wrt $X$ localizes the integral on the boundary of the 4-simplex defined by the product of theta functions and delta function. This together with the fact that $P_5\cdot W_i=\delta_{5i}$ turns the integral into one on the tetrahedron.

If we now focus on the full integral and not just on the integrand, we have
	\bea
	&&\int \ud Q\,\int_{\mathds R^D} \ud X e^{Q\cdot X}\, P_5\cdot(-\partial_X) \frac{\prod_{i=1}^5 \Theta(W_i\cdot X) \delta(W_6\cdot X)}{\sqrt{\det P_i\cdot P_j}}=\nonumber \\
	&&\int e^{X^2} P_5\cdot(-\partial_X) \prod_{i=1}^5 \frac{\Theta(W_i\cdot X) \delta(W_6\cdot X)}{\sqrt{\det P_i\cdot P_j}}
	=(\det P_i\cdot P_j)^{-\frac 12}\int_{\mathcal P_5}\ud X\, (P_5\cdot X)
	\eea
where in the last step we have integrated by parts and done the integral over the scale $|X|$. In this last form, the integral is over a 4-simplex $\mathcal P_5$ in $AdS_5$. And yet, we know that the very same integral is given by the volume of the tetrahedron with vertices given by the first four $P_i$. In this way we have derived a non-trivial relation between a moment of a 4-simplex and the volume of one of its faces.

Now let us consider the slightly different integral,
	\bea
	I_5=\int \ud Q\, \frac{Y\cdot Q}{\prod_{i=1}^5 P_i\cdot Q}
	\eea
We will impose special kinematics: we demand that $P_i\cdot P_{i+1}=0$, that is, the coordinates $x_i$ and $x_{i+1}$ are null separated. Further for $Y$ we shall take the special vector
	\bea
	Y=-P_{24} P_{35} P_1+P_{14}P_{35} P_2+P_{14}P_{25} P_3-P_{13}P_{25} P_4+P_{13}P_{24} P_5
	\eea
It is clear that with these special choices, the five-point integral above collapses into a sum of five appropriately normalized box integrals.  The normalization has been chosen such that each of these box integrals yields the volume of a tetrahedron (i.e. the $\sqrt{\det P_{ij}}$ is taken out). There is another consequence of these choices: $Y\cdot P_i=0$ for all $i\neq 5$. Thus the integral above is closely reminiscent of the chiral pentagon integral \cite{ArkaniHamed:2010gh}. 

Its calculation is exactly the same as in our previous example and so we find
	\bea
	I_5=\int e^{X^2} \,Y\cdot(-\partial_X) \prod_{i=1}^5 \frac{\Theta(W_i\cdot X) \delta(W_6\cdot X)}{\sqrt{\det P_i\cdot P_j}} \nonumber \\
	=\int e^{X^2} \,(Y\cdot X) \prod_{i=1}^5 \frac{\Theta(W_i\cdot X) \delta(W_6\cdot X)}{\sqrt{\det P_i\cdot P_j}}
	\eea
So, this kind of integrals is associated with the computation of a certain moment of a $4$-simplex, the moment being calculated along the direction $Y$.

This last form can be rewritten again. Indeed it is clear that due to our special kinematics, $Y\cdot X$ is only non-zero if $X$ contains a piece proportional to $P_5$. Such a piece can be captured by taking its internal product with $W_5$, and so we also have
	\bea
	I_5=\frac{Y\cdot P_5}{\sqrt{\det P_i\cdot P_j}} \int_\mathcal P \ud X\, (W_5\cdot X)
	\eea
This result holds for any $Y$ so far as $Y\cdot P_i=0$ for all $i<5$. In particular this shows that the choice of $Y$ in \cite{ArkaniHamed:2010gh} and ours lead to completely equivalent results: their $Y$ differs from ours by an additional piece $\simeq \epsilon P_1 P_2 P_3 P_4 P_5$, the coefficient of which is chosen such that $Y^2=0$. Since this additional piece vanishes contracted into $P_5$, we see that both choices are equivalent. Incidentally this shows that in this simple 5-point example the chirality of the chiral pentagon is actually completely irrelevant.

The chiral pentagon integral, due to the presence of its special numerator factor, actually reduces to a sum of boxes, and this can be seen directly in the spline picture. Previously it had been found that the $n$-pt 1-loop MHV amplitudes correspond geometrically to the volume of the boundaries of closed polytopes. The $5$ pt case is an especially simple case of this, where this closed polytope is a 4-simplex; 4-simplices are associated with integrands with 5 propagators, and the presence of a numerator naturally gives the boundary of this 4-simplex. In order for this to happen, we would need a numerator of the form $Y'\cdot Q$, with
	\bea
	Y'=P_{24} P_{35} P_1+P_{14}P_{35} P_2+P_{14}P_{25} P_3+P_{13}P_{25} P_4+P_{13}P_{24} P_5
	\eea
which we see differs from $Y$ by the presence of two single-mass box integrals. This of course reproduces the MHV amplitude representation of \cite{ArkaniHamed:2010gh} as it should.

Similar arguments could be made for 6-pt amplitudes in terms of a ``chiral hexagon'', and we suspect that a slightly more complicated reasoning may well be able to shed light on the simple integral representations of \cite{ArkaniHamed:2010gh} for higher point functions. 

\section{Simplicial decompositions}
\label{simplicial}

\subsection{Method 1: Partial Fractions}
When the number of nodes is larger than $D$, interesting combinatorial structures appear; the polyhedron becomes divided into simplicial cells across which the spline has discontinuous derivatives. These phenomena have been systematically studied in the case $\Delta_i=1$, less so otherwise. In this section we focus on this simple case, although some of our results can be easily generalized.

Recall that the definition of the $D$-dimensional spline is
	\bea
	\mathcal T(X;\{P_i\})=\int_0^{+\infty} \ud t\, \delta^{(D)}(X-\sum t_i P_i)
	\eea
We can think of the $\delta$-function as imposing a set of linear equations in $t$-space,
	\bea
	\Bigg(P_1\Bigg| \ldots \Bigg |P_n\Bigg)\cdot
	\left(
	\begin{tabular}{c}
	$t_1$ \\
	\vdots\\
	$t_n$
	\end{tabular}
	\right)=\Bigg(X\Bigg).
	\eea
These constraints, together with the positivity conditions $t_i>0$ define a convex polytope in $t$-space. In fact, any convex polytope can be defined by such equations. It is usually called a variable polytope, since its shape depends on the vector $X$. We emphasize that this polytope does not have to be a simplex. Also that this polytope is completely different from the hyperbolic simplices we have seen before - the latter lived in ``target-space'', while the former lives in $t$-space.

To be more precise, the polytope can only exist if $n$, the number of nodes, is larger than $D$, otherwise the spline is a tempered distribution. For $n=D$ the polytope reduces to a point, for $n=D+1$ it is generically a line, and so on. In any case, it is clear that the spline simply computes the $n-D$ dimensional volume of this polytope, up to a prefactor. In particular, for $n=D$, the volume is equal to one as long as $X$ lies in the polyhedral cone defined by the nodes $P$, and we recover our previous formula. For $n>D$ we expect that the volume should scale like $|X|^{n-D}$, and so the spline must be given by a polynomial homogeneous in $X$ of that degree. It is clear however that there should be more interesting structure regarding the dependence on $X$. Indeed, at special lower codimension surfaces in $X$-space the $t$-polytope can degenerate onto lower dimensional ones. Across these surfaces the homogeneous polynomials have to be continuous, but their derivatives do not. In target space, we have a polyhedral cone defined by the nodes $P$, and this cone is generically not simplicial for $n>D$. However, it can be decomposed as a sum of simplicial cones, which define a cell structure on the original cone. As $X$ moves from one cell to the next, the derivatives of the polynomials can jump.

\begin{figure}
	\centering
	\includegraphics[scale=0.5]{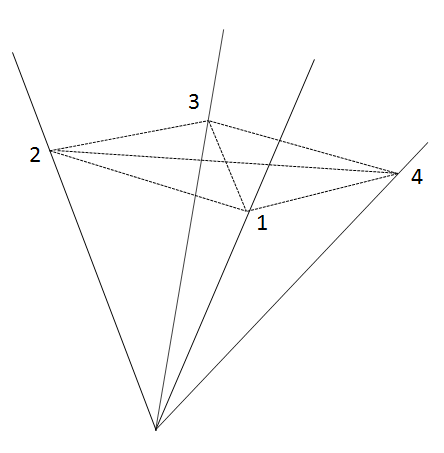}
	\caption{A 4 vertex cone and its decomposition into simplicial 3-vertex ones. One needs three of these to obtain the same cell structure.}
	\label{fig:Cone4pt}
\end{figure}

To analyse this structure, and determine precisely the form of the spline, we can use the Laplace transform, or in other words, go to the integrand form that we know and love. There we can use elementary linear algebra to simplify the integrand and transform back to figure out what is happening geometrically. We begin with the scalar box integrand, which is the Laplace transform of a four-spline. But now, we imagine choosing kinematics such that the nodes $P_i$ lie on a 3 dimensional sub-space. What this means is that we are picking all the $x_i$ coordinates to lie on a one-dimensional subspace.
To see that something must occur, recall that the scalar box {\em integral} is equal to the volume of a tetrahedron in hyperbolic space. If we demand that our kinematics lie on a one-dimensional subspace, our tetrahedron gets squashed, because it cannot live on an $AdS_2$. This is shown geometrically in figure \ref{fig:Cone4pt}.
Accordingly, the box integral naively vanishes. However, we are forgetting that actually the unnormalized box integral is given by a volume {\em ratio}:
	\bea
	{\hbox{\lower 12pt\hbox{
\begin{picture}(30,30)
\put(0,0){\includegraphics[width=30pt]{1loopmassive.pdf}}
\end{picture}
}}}=\frac{\mbox{vol $\mathcal P_4$}}{\sqrt{\det P_{ij}}}
	\eea
The denominator must also vanish for 1d kinematics, since it is one of the Gram determinants, and hence we may hope for a finite final answer. This is indeed the case, as can be seen from the explicit expression for the box integral:
	\bea
	{\hbox{\lower 12pt\hbox{
\begin{picture}(30,30)
\put(0,0){\includegraphics[width=30pt]{1loopmassive.pdf}}
\end{picture}
}}}=\frac{2 \mbox{Li}_2(-u \rho)+2 \mbox{Li}_2(-v \rho)+\log\left(\frac vu\right)\log\left(\frac{1+\rho v}{1+\rho u}\right)+\log(\rho u) \log(\rho v)+\frac{\pi^2}3}{\sqrt{\Delta}}\, 
	\eea
with
	\bea
	\rho=\frac{2}{1-u-v+\sqrt{\Delta}}, \qquad \Delta=(1-u-v)^2-4 u v.
	\eea
If we now set $u=z^2, v=(1-z)^2+\epsilon$ and expand for $\epsilon\ll 1$, we find that $\Delta\simeq \epsilon$, and so the numerator of the box integral must be vanishing as well. This is of course nothing but the hyperbolic volume going to zero. The resulting expression is
	\bea
-\frac{\log(1-z)-\log\left(\frac{1}{1-z}\right)}{2z(1-z)}-\frac{\log(z)-\log\left(\frac{1}{z}\right)}{2 z(1-z)}+\frac{\log\left(\frac{(1-z)^2}{z^2}\right)}{1-z}-\frac{\log\left(\frac{(1-z)^2}{z^2}\right)}{2z(1-z)} \label{boxlog}
	\eea
Depending on the kinematic range - whether $z$ is negative, between $0$ and $1$ or larger than one - the expression can simplify. Now let us try to interpret what's going on in the spline picture. We start off with the box integrand
	\bea
	\frac{1}{(P_1\cdot Q)(P_2\cdot Q)(P_3\cdot Q)(P_4\cdot Q)}
	\eea
Since we are in $1d$ kinematics, there can be at most three independent $P$ vectors (since they live in the $d+2$ dimensional embedding space). Define the vectors orthogonal to the $P_i$, i.e. $W^1_{23}\cdot P_1=1, W^1_{23}\cdot P_{2,3}=0$. Then we have
	\bea
\prod_{i=1}^4 \frac{1}{P_i\cdot Q}&=&
\frac{W^1_{23}\cdot P_4}{(P_2\cdot Q)(P_3\cdot Q)(P_4\cdot Q)^2}+
\frac{W^2_{31}\cdot P_4}{(P_1\cdot Q)(P_3\cdot Q)(P_4\cdot Q)^2}+
\frac{W^3_{12}\cdot P_4}{(P_2\cdot Q)(P_2\cdot Q)(P_4\cdot Q)^2}
\eea
Hence the box integral has decomposed into a sum of triangle integrals, which can be easily done via the star triangle relation. Geometrically, the 4-vertex cone is being decomposed as a sum of three simplicial cones; algebraically the 4-spline is written as a sum of three 3-splines. Consider for instance the first term:
	\bea
\frac{W^1_{23}\cdot P_4}{(P_2\cdot Q)(P_3\cdot Q)(P_4\cdot Q)^2}&=& W^1_{23}\cdot \partial_Q\left[ \frac{1}{(P_2\cdot Q)(P_3\cdot Q)(P_4\cdot Q)}\right]\nonumber \\
&& =\int \ud X\, e^{-Q\cdot X} (W^1_{23}\cdot X)\, \mathcal T_{1,1,1}(X,\{P_2,P_3,P_4\})
\eea
So we have a polynomial, $W^1_{23}\cdot X$ living on a triangular cell. The full 4-spline can hence be written as the sum of three pieces, corresponding to three-overlapping simplicial cones, which put together make up the polyhedral cone with vertices $P_1,P_2,P_3,P_4$. On each piece there lives a different linear polynomial in $X$.

This basic example can be easily generalized to a higher number of nodes or higher dimensionality and the conclusions are similar. So much for the computation of the spline, and hence for the integrand. But what does this mean for the {\em integral}?

If we do the $Q$ integral something very interesting happens. The $e^{-Q\cdot X}$ becomes $e^{X^2}$, and hence the polynomial can be written as a derivative of that factor
	\bea
	&&\int \ud X e^{X^2}(W^1_{23}\cdot X)\, \mathcal T_{1,1,1}(X,\{P_2,P_3,P_4\})=\int \ud X e^{X^2}(W^1_{23}\cdot \partial_X^{\leftarrow})\, \mathcal T_{1,1,1}(X,\{P_2,P_3,P_4\}) \nonumber \\
	&&=-\int \ud X e^{X^2}(W^1_{23}\cdot \partial_X^{\rightarrow})\, \mathcal T_{1,1,1}(X,\{P_2,P_3,P_4\}).
	\eea
Acting on the spline we pick up the boundary of the region we are working in, so that the integral further localizes on the lines that make up the triangular cell! Since all the weights are one, we will end up getting lengths of lines in $AdS_2$ which are however divergent. Such a divergence can also be seen directly from the point of view of the triangle. The dimensions $2,1,1$ lead to a gamma function with zero argument, and a $P_{ij}$ with zero exponent. To regulate we can set $P_i^2=\epsilon$ and send $\epsilon$ to zero at the end of the calculation. The length of a line in $AdS$ goes like a $\log$ so we're clearly on the right track to recover \reef{boxlog}. More concretely, denoting the length of the line between points $i$ and $j$ by $L(i,j)$ we have
	\bea
L(i,j)=\frac{\log\left(\frac{P_{ij}^2}{P_{ii} P_{jj}}\right)}{\sqrt{P_{ij}^2-P_{ii} P_{jj}}}
	\eea
and
	\bea
{\hbox{\lower 12pt\hbox{
\begin{picture}(30,30)
\put(0,0){\includegraphics[width=30pt]{1loopmassive.pdf}}
\end{picture}
}}}\Bigg |_{\mbox{\tiny 1d}}&=&(W^{1}_{23}\cdot W^{2}_{34})\, L(3,4)+(W^{1}_{23}\cdot W^{3}_{42})\, L(2,4)+(W^{1}_{23}\cdot W^{4}_{23}) \, L(2,3)\nonumber \\
	&+&(W^{2}_{31}\cdot W^{1}_{34})\, L(3,4)+
	(W^{2}_{31}\cdot W^{3}_{41})\, L(1,4)+
	(W^{2}_{31}\cdot W^{4}_{13})\, L(1,3)\nonumber \\
	&+&(W^{3}_{12}\cdot W^{1}_{24})\, L(2,4)+
	(W^{3}_{12}\cdot W^{2}_{41})\, L(1,4)+
	(W^{3}_{12}\cdot W^{4}_{12})\, L(1,2)
	\eea	
with {\em e.g.}
	\bea
	W^{1}_{23}\cdot W^{2}_{34}=\frac{\delta^{AB}_{CD}\, P_2^C\, P_3^D\, P_{3,A}\, P_{4,B}}{\delta^{ABC}_{DEF} \,P_1^D\, P_2^E\, P_3^F\, P_{2,A}\, P_{3,B}\, P_{4,C}}.
	\eea
A straightforward calculation now leads precisely to result \reef{boxlog}.

Although our calculations were performed for the scalar box, it is clear that the method is generically applicable to higher point integrals. By applying the partial fraction decomposition we can successively reduce any product of propagators to products involving only $D$ of them. Since for such products we already know the spline representation, we are done. Geometrically the polyhedron gets decomposed into simplices, whose overlap reproduce the ``big-cell'' \cite{hyperplane} decomposition of the original polyhedron. At least this is all true for the case where all weights equal one. To the best of our knowledge the generic case has not been addressed in the literature, and it indeed it seems hard to determine the spline in those cases, since the partial fraction certainly fails if the weights are not integers. Next we describe another method for determining the spline decomposition, which is also only valid in the unity weights case.
	
\subsection{Method 2: Residue computations}

We have already seen the method of partial fractions in action in the decomposition of the box into lines. As it turns out there is a more efficient way of doing these computations in terms of {\em residues}.

Let us go back to figure \ref{fig:Cone4pt}, showing our squashed tetrahedron. Although we see from the picture that there are several possible simplicial cones, there is a minimal (non-unique) set which paves the entire cone. In this case it is easy to check that there are only three triangles required. Algebraically, we have a set of vectors which are not all linearly independent, and from these we want to look at the set of possible basis. To each basis corresponds a single simplicial cone. The question is then: what is a minimal set of basis one must use to pave the full cone? The answer is the set of {\em unbroken basis}. This is a combinatorial concept which can be understood as follows. Consider an ordered set of vectors $P_i$, and choose a certain basis labelled by integers $i_1<i_2<\ldots<i_D$. We say that the basis is {\em broken} if there is a vector $P_j$ such that $j<i_D$ and $P_j$ is linearly dependent on the vectors $i_k$ with $k>j$. In other words, if this vector can be written as a linear combination of the basis elements to its right on the list. With this definition, it is clear that any unbroken basis must include the very first vector. 

In practice, if we assume generic kinematics, finding the set of unbroken basis is very simple. One fixes the first vector which has to be $P_1$, and then chooses all possible $D-1$ subsets of the remaining $n-1$ vectors. So for instance, in the four point case given above, a set of unbroken basis is $(1,2,3),(1,2,4),(1,3,4)$, and it is easily checked that the corresponding simplicial cones cover the full four-point polyhedral cone. In general we therefore expect there to be $(n-1)!/(D-1!)(n-D)!$  unbroken basis for generic kinematics. So, for a 6-node spline in $D=4$ (in other words, $2d$ kinematics since the embedding space is $D=d+2$) we expect a paving of the hexahedral cone by 10 tetrahedral cones, corresponding to the unbroken bases $(1,2,3,4),(1,2,3,5),\ldots,(1,4,5,6)$.

Going back to the spline, we can write it as the sum of several pieces, each corresponding to one of the elementary simplicial cones. On each of these the spline will be given by a polynomial, and it is therefore our next task to determine the polynomial associated to each unbroken basis. To determine the polynomial we compute a residue, and this is done as follows: if the elements of the unbroken basis are $P_{i_k}$ we do the map
	\bea
	P_{i_1}\cdot Q\to u_1, \quad, P_{i_2}\cdot Q\to u_1 u_2, \quad \mbox{etc}.
	\eea
We now take the Laplace transform of the spline, which is $L\equiv\prod_{i=1}^n (P_i\cdot Q)^{-1}$, do this map and multiply by the form 
\bea
e^{X\cdot Q}\ud P_{i_1}\cdot Q \wedge \ud P_{i_2}\cdot Q \wedge \ldots=e^{X\cdot Q} u_1^{D-1}u_2^{D-2}\ldots u_{D-1} \ud u_1\wedge \ud u_2\wedge\ldots \wedge \ud u_D
\eea
In the above we can always decompose $X$ as a sum of $P_i$ so the exponential is also a function of the $u_i$. We think of these $u_i$ as complex variables and compute the multidimensional residue. This is quite interesting, since we'll get an expression of the form
	\bea
	\int \prod_i \frac{\ud u_i}{u_i} f(u_i)
	\eea
with $f(u_i)$ an analytic function. In terms of the original variables notice the measure is
	\bea
	\int \ud \log(P_1\cdot Q)\, \ud \log\left(\frac{P_2\cdot Q}{P_1\cdot Q}\right)\,\ud \log\left(\frac{P_3\cdot Q}{P_2\cdot Q}\right) \ldots \ud \log\left(\frac{P_D\cdot Q}{P_{D-1}\cdot Q}\right)
	\eea
Let us consider an  example, for instance when $n=D+1$, and let us consider the no-broken basis $b$ formed by the first $D$ elements. Then we get that %
	\bea
	&&L e^{X\cdot Q} u_1^{s-1}u_2^{s-2}\ldots u_{s-1} \ud u_1\wedge \ud u_2\wedge\ldots \wedge u_s=\nonumber \\ &&\frac{\ud u_1\wedge \ud u_2\wedge\ldots \wedge u_s}{u_1\ldots u_s}\,\frac{1+W^{(b)}_1\cdot X\, u_1 +W^{(b)}_2\cdot X\, u_1 u_2+ \ldots}{u_1(W^{(b)}_1\cdot P_n+u_2 W^{(b)}_2\cdot P_n+\ldots)}
	\eea
Here $W^{(b)}_i$ are the vectors satisfying $W^{(b)}_i\cdot P^{(b)}_j=\delta_{i,j}$.	
The residue is simply
	\bea
	\frac{W_1^{(b)}\cdot X}{W_1^{(b)}\cdot P_n}.
	\eea
Now let us consider the slightly more  complicated example $n=D+2$. For each unbroken $b$ let us denote $\hat P_1^{(b)}, \hat P_2^{(b)}$ the two elements which are not in it. Then it is easy to check that the polynomial corresponding to a given basis $b$ is simply
	\bea
	\frac{(W_1^{(b)}\cdot X)^2}{(W_1^{(b)}\cdot \hat P_1^{(b)})(W_1^{(b)}\cdot \hat P_2^{(b)})}
	\eea
This means that the full spline can be written in this case as
	\bea
	T(X;\{P_i\})=\sum_{\{b\}} \frac{(W_1^{(b)}\cdot X)^2}{(W_1^{(b)}\cdot \hat P_1^{(b)})(W_1^{(b)}\cdot \hat P_2^{(b)})}\, \frac{\chi_{(b)}}{\sqrt{\det b^T\,b}}
	\eea
where $\chi_{(b)}$ is the characteristic function of the simplicial cone corresponding to the basis $b$, namely
	\bea
	\chi_{(b)}=\prod_{i=1}^D \Theta(W_i\cdot X)
	\eea
and $\det b^T\,b$ is the determinant of the matrix of the $P_{ij}$, where the $i,j$ run over the elements of the basis $b$.

Again, we have illustrated the residue method with these simple examples, but more general cases are also easily determined. In the non-generic case where there are extra linear deneracies the algorithm is unmodified; one must simply be more careful in constructing the set of unbroken basis.

\subsection{Doing the integrals}

Now we come to the interesting part. We are not interested in the spline, but rather in its integral over a gaussian measure,
	\bea
	I=\int \ud X e^{X^2} \mathcal T(X;\{P_i\}).
	\eea
Let us consider the case $n=D+2$. Then the integral above is a sum of several pieces, each of which takes the form
	\bea
	\int \ud X e^{X^2} \frac{(W_1^{(b)}\cdot X)^2}{(W_1^{(b)}\cdot \hat P_1^{(b)})(W_1^{(b)}\cdot \hat P_2^{(b)})}\, \frac{\chi_{(b)}}{\sqrt{\det b^T\,b}}
	\eea
But now we see something very interesting: because of the presence of the gaussian measure, we can write the polynomials as derivatives, and integrate by parts! There are then two kinds of terms that appear. One of them is simply
\bea
\frac{(W_1^{(b)})^2}{(W_1^{(b)}\cdot \hat P_1^{(b)})(W_1^{(b)}\cdot \hat P_2^{(b)})} \int \ud X e^{X^2} \, \frac{\chi_{(b)}}{\sqrt{\det b^T\,b}} \label{trace}
\eea
which is, up to an overall factor, the volume of the $D-1$-dimensional simplex in $AdS$ with vertices given by the unbroken basis $b$. The other kind of terms correspond to lower dimensional simplices, which arise by hitting the characteristic function with derivatives. Now, the remarkable thing is that if the all simplices are ideal (which means that all the $P_i$ square to zero\footnote{In terms of loop integrals, this means that all propagators are those of massless particles.}), these terms seem to cancel among themselves! Although we have no formal proof, we have checked in several examples that this is the case. One way of understanding this might be that such terms would lead to integrals with necessarily lower transcendentality than those such as \ref{trace}.

To summarize, we conjecture the following: the $d$-dimensional $d$-gon loop-integral can be written, upon using $d-4$ dimensional kinematics as a sum of $(d-2)$ dimensional $d-2$-gon loop integrals. In particular this implies that an hexagon integral in $2d$ kinematics can be written as a sum of $10$ box integrals; and that an octagon integral in $4d$ kinematics can be written as a sum of $7!/5!2!=21$ hexagon integrals. Each integral is labelled by an unbroken basis containing $d-2$ elements, and the coefficient of each such integral is
	\bea
	\frac{(W_1^{(b)})^2}{(W_1^{(b)}\cdot \hat P_1^{(b)})(W_1^{(b)}\cdot \hat P_2^{(b)})}
	\eea
This is only one of many other possible conjectures which would hold for more complicated integrals with $n>D+2$. Work is in progress on these matters \cite{workinprogress}. 

\section{Spline convolutions and multiloop results}
\label{multiloop}
\subsection{Splitting the tetrahedron}
Let us go back one last time to the four-dimensional scalar box integral. Thanks to the spline, it can be nicely interpreted as the 3-volume of a hyperbolic tetrahedron (up to a normalization factor). Now, it is clear that at least in Euclidean space, one can also see a 3-volume as roughly the product of a length by an area; or as a triple product of lengths. So is there an analogous statement in hyperbolic space? We shall now see that indeed there is, and the answer is connected to the circumstances under which splines can be merged together. Indeed, let us again consider the box integrand:
\bea
\frac{1}{(P_1\cdot Q)\, (P_2 \cdot Q)\, (P_3 \cdot Q)\, (P_4 \cdot Q)}
\eea
The connection of this integrand with the four-noded spline is directly related to the procedure of joining the four separate propagators into a single one by use of Feynman parameters. But this is not the only choice: we could for instance join three of them, or make two groups of two. This leads to the appearance of several splines, which however must be exactly equivalent to the original one. We can see this quite directly. Recall that we have found
	\bea
	{\hbox{\lower 12pt\hbox{
\begin{picture}(30,30)
\put(0,0){\includegraphics[width=30pt]{1loopmassive.pdf}}
\end{picture}
}}}=\iads \ud X \mathcal T(X,\{P_i,\Delta_i\})=\iads \ud X \int_0^{+\infty} \prod_{i=1}^4 \ud t_i\, \delta(X-\sum t_i P_i).
	\eea
Now we can repeat our original trick of introducing delta functions. For instance,
	\bea
	&&\int_{\mathds M^D}\!\! \ud X_1 \ud X_2\,\iads \!\ud X \int_0^{+\infty}\! \prod_{i=1}^4 \ud t_i\, \delta(X\!-\!X_1\!-\!X_2)\nonumber \\
	&& \hspace{2 cm} \delta(X_1\!-\!t_1 P_1\!-\!t_2 P_2)\,\delta(X_2\!-\!t_3 P_3\!-\!t_4 P_4).
	\eea
To continue, let us separate the integrals over the scale factors $|X_1|,|X_2|$ and denote these two variables by $s_1, s_2$. Then it is easy to find
	\bea
	{\hbox{\lower 12pt\hbox{
\begin{picture}(30,30)
\put(0,0){\includegraphics[width=30pt]{1loopmassive.pdf}}
\end{picture}
}}}=&&\iads \ud X \ud X_1 \ud X_2\,  \int_0^{+\infty}  \frac{\ud s_1}{s_1}\frac{\ud s_2}{s_2} s_1^{2}  s_2^{2}\, \delta(X-s_1 X_1-s_2 X_2)\times \, \nonumber \\
	&& \int_0^{+\infty} \ud t_1 \ud t_2\, \delta(X_1-t_1 P_1-t_2 P_2)\, \int_0^{+\infty} \ud t_3 \ud t_4\, \delta(X_2-t_3 P_3-t_4 P_4)\nonumber \\
=&& \iads \ud X \ud X_1 \ud X_2\, \mathcal T_{2,2}(X,\{X_1,X_2\})\, \mathcal T(X_1,\{P_1,P_2\})\mathcal T(X_2,\{P_3,P_4\})
	\eea
Written in this form this might seem a bit obtuse, but the interpretation is actually quite simple. Firstly, there are the two splines with arguments $X_1$, $X_2$. These have support on the hyperbolic geodesic lines from $P_1$ to $P_2$ and $P_3$ to $P_4$ respectively. The remaining spline has support on the hyperbolic geodesic connecting $X_1$ and $X_2$ itself. Notice though that the weights of the nodes are no longer $1$. This seems to be the closest one can get to a natural splitting of the volume into three ``lengths''. The use of inverted commas is very important here, since as we've shown before, the integral over the spline gives at best a volume {\em ratio}, and only in the case where all the weights are equal to one. In this case we see that there is no getting around the fact that the spline is a ratio; no factor can be multiplied to cancel the denominator of the spline with argument $X$, since such a denominator involves $X_1, X_2$ which are being integrated over.

It is also easy to show that there is an alternative representation in terms of only two splines,
	\bea
	{\hbox{\lower 12pt\hbox{
\begin{picture}(30,30)
\put(0,0){\includegraphics[width=30pt]{1loopmassive.pdf}}
\end{picture}
}}}&=&\iads \ud X \ud X'\, \mathcal T_{1,3}(X,\{P_4,X'\})\, \mathcal T(X',\{P_1,P_2,P_3\}) \nonumber \\
	&&=\iads \ud X\frac{1}{(P_4\cdot X)^3}\, \mathcal T(X,\{P_1,P_2,P_3\}) \label{twospline}
	\eea
Notice the appearance of $1/(P_4\cdot X)$, which is nothing but the scalar bulk to boundary propagator in $AdS$! Such a factor is precisely the $AdS$ integral of the spline with nodes $P_4,X$. Since the bulk-to-boundary propagator is a limit of the bulk-to-bulk propagator, one may naturally wonder what the result would be if instead of the massless $P_4$ we had a massive, $AdS$ vector. We show in the appendix that $\iads \ud X\,  \mathcal T_{\Delta_1,\Delta_2}(X,\{X_1,X_2,\})$ is essentially the sum of two $AdS$ bulk to bulk scalar propagators. In the case where $\Delta_1=\Delta_2=2=d/2$ one obtains the logarithmic bulk-to-bulk propagator of a field of dimension $d/2$ in $AdS$. The conclusion is that the original flat space integral is given entirely in terms of natural quantities in $AdS$ space.

There are several obvious generalizations of these results. The number of nodes involved and their type (massive or not), the dimensionality of spacetime, the weights and so on are all quite irrelevant. In this way, e.g. a hexagon integral can be written in a number of different ways: as the volume of a hyperbolic 5-simplex; two hyperbolic triangles glued by a line; a triangle with one of its tips lying inside a tetrahedron; and so forth. The rule is that if a given node $X'$ of a spline $\mathcal T'$ is the argument of a spline $\mathcal T''$, and if this argument $X'$ is being integrated over, then both splines can be merged into a single one as long as the sum of the weights in $\mathcal T''$ is equal to the weight of the node $X'$ in $\mathcal T'$. So, for instance
	\bea
	\iads \ud X'\, \mathcal T_{1,3}(X,\{P_4,X'\})\, \mathcal T(X',\{P_1,P_2,P_3\})=\mathcal T(X,\{P_1,P_2,P_3,P_4\}).
	\eea
These rules are useful since sometimes a given representation in terms of some number of splines might be easier to derive than others, especially when one is considering multi-loop integrals. It is to these that we turn next.

\subsection{Multiloops}
For definiteness, we shall start by considering the fully-massive double-box integral, which in the dual conformal picture corresponds to a 3-to-3 tree-level exchange diagram. This is displayed in figure \ref{fig:2loop}. The integral is written
\bea
{\hbox{\lower 14pt\hbox{
\begin{picture}(40,37)
\put(0,0){\includegraphics[width=40pt]{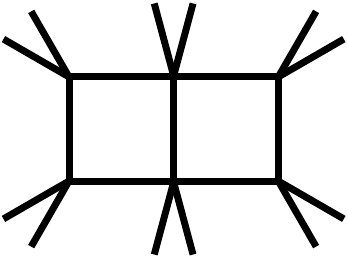}}
\end{picture}
}}}=\int \frac{\ud Q_1 \ud Q_2}{(2\pi^{d/2})^2}\, \frac{1}{Q_1\cdot Q_2}\, \prod_{i\in L} \frac{1}{P_i\cdot Q_1}\, \prod_{i\in R} \frac{1}{P_i\cdot Q_2}
\eea
Our first question is: can such a integral be written solely in terms of $AdS$ quantities, or was this rewriting a lucky accident at one-loop? As usual, introduce Schwinger parameters
	\bea
	\int \frac{\ud Q_1 \ud Q_2}{(2\pi^{d/2})^2}\,\int_0^{+\infty} \ud s\,\prod_{i\in L} \ud t_i\,\prod_{j\in R} \ud t_j \, e^{Q_1\cdot(\sum_{i\in L} t_i P_i)+Q_2\cdot(\sum_{i\in R} t_i P_i)+s Q_1\cdot Q_2}
	\eea
There are several ways to proceed now. We could introduce splines, one for each side of the diagram. This would correspond to two triangles connected by a convolving kernel which turns out to be the integral of a 2-spline. Instead, we first do the $Q_1$ integral, giving
	\bea
	\int \frac{\ud Q_2}{(2\pi^{d/2})}\,\int_0^{+\infty} \ud s\,\prod_{i\in L} \ud t_i\,\prod_{j\in R} \ud t_j \, e^{ (\sum_{i\in L}t_i P_i)^2+ Q_2\cdot[s (\sum_{i\in L}t_i P_i)+\sum_{i\in R} t_i P_i]}
	\eea
Performing the $Q_2$ integral we can write the result in turns of two splines,
	\bea
	\int_{\mathds M^6\times \mathds M^6}\!\!\! \ud X_1 \ud X_2\, e^{X_1^2+X_2^2} \mathcal T(X_1,P_L)\, \mathcal T(X_2,\{X_1,P_R\}).
	\eea
with $P_{L,R}\equiv \{P_i, i\in L,R\}$. Homogeneity in the norms of $X_1$ and $X_2$ allows us to perform these integrals easily, and we find
	\bea
	{\hbox{\lower 14pt\hbox{
\begin{picture}(40,37)
\put(0,0){\includegraphics[width=40pt]{2loopmassive.pdf}}
\end{picture}
}}}=\iads\!\! \ud X_1 \ud X_2\, \mathcal T(X_1,P_L)\, \mathcal T(X_2,\{X_1,P_R\})
	\eea
The interpretation is clear: we are doing an integral over a tetrahedron with vertices $X_1,P_4,P_5$ and $P_6$; at the same time we are integrating over the position of vertex $X_1$, but not over all of $AdS$, rather inside the hyperbolic triangle with vertices $P_1,P_2,P_3$.

\begin{figure}
	\centering
		\includegraphics[scale=0.5]{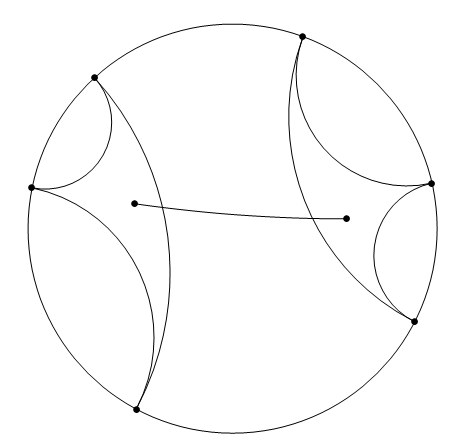}
	\caption{An $AdS$ picture of the double box integral.}
	\label{fig:3to3}	
\end{figure}

Using our merger rules, we can derive the alternative representation 
	\bea
	{\hbox{\lower 14pt\hbox{
\begin{picture}(40,37)
\put(0,0){\includegraphics[width=40pt]{2loopmassive.pdf}}
\end{picture}
}}}=\iads \prod_{i=1}^3\ud X_i\,\, \mathcal T(X_1,P_L)\, \mathcal T_{1,3}(X_2,\{X_1,X_3\})\, \mathcal T(X_3,P_R).
	\eea
which can be visualized geometrically in figure \ref{fig:3to3}.
We have split the tetrahedron into a line and a triangle, so that the very same integral is seen as two hyperbolic triangles connected by a line. More precisely, this line is precisely the two-spline mentioned in the previous section, and computed more carefully in the appendix, corresponding to the sum of two bulk-to-bulk scalar propagators in $AdS$; one each of dimensions $3$ and $1(=d-3)$. We could of course keep going and further split the triangle into a pair of lines. However, it seems impossible to join the tetrahedron and triangle into a single object.

Of course, everything we have said up to now is easily generalizable to more complicated integrals. Weights can be varied, masses can be added, and so on. Once this is done, another kind of phenomenon appears. Consider a $2\to 2$ exchange diagram with given weights $\Delta_i$ and an internal propagator of weight $\delta$. Then it is not hard to show that this diagram is given by

	\bea
	&&\gt{\Delta_1+\Delta_2-\delta}\int \prod_{i=1}^3 \ud X_i~ \mathcal T_{\Delta_1,\Delta_2}(P_1,P_2; X_1)\nonumber \\
	&& \hspace{3 cm}\, \mathcal T_{\delta,\Sigma-\delta}(X_1,X_2; X_3)\,
\mathcal T_{\Delta_{3},\Delta_{4}}(P_3,P_4; X_2).
	\eea
This is quite interesting, because the $4$-pt contact interaction admits a very similar representation, once we split the tetrahedron onto a product of three lines:
	\bea
	\int \prod_{i=1}^3 \ud X_i~ \mathcal T_{\Delta_1,\Delta_2}(P_1,P_2; X_1)\, \mathcal T_{\Delta_1+\Delta_2,\Delta_3+\Delta_4}(X_1,X_2; X_3)\,
\mathcal T_{\Delta_{3},\Delta_{4}}(P_3,P_4; X_2).
	\eea
Equality can be only be achieved if $\Delta_1+\Delta_2=\delta$, at which point the exchange diagram develops a pole from the gamma function. But so, we have learned that by looking at the behaviour of integrals when the {\em weights} are varied, we can find new connections between integrals with different numbers of loops. In the case in point, we see that up to a controllable divergent pre-factor, the one-loop integral is a special case of the two-loop one.

\begin{figure}
	\centering
		\includegraphics[scale=0.5]{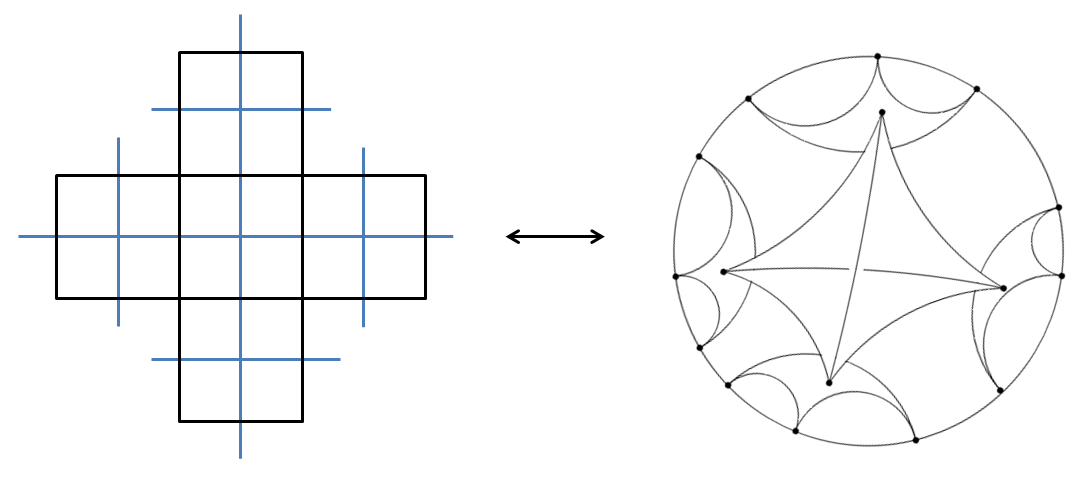}
	\caption{A (fully massive) 4-loop integral, its position space dual and its glued up version in $AdS$ space.}
	\label{fig:12pt2}
\end{figure}

Let us close this section by a few further comments. This two-loop example shows that the generalized star-triangle relations continue to hold beyond one-loop. Stars get glued up onto hyperbolic simplices, and flat-space integrals transform into $AdS$ ones. For instance, the symmetric 12-pt diagram in $\phi^4$ theory gets beautifully glued up onto a picture of a hyperbolic tetrahedron whose vertices lie on  ideal hyperbolic triangles, as shown in figure \ref{fig:12pt2}. 

\begin{figure}
	\centering
		\includegraphics[scale=0.7]{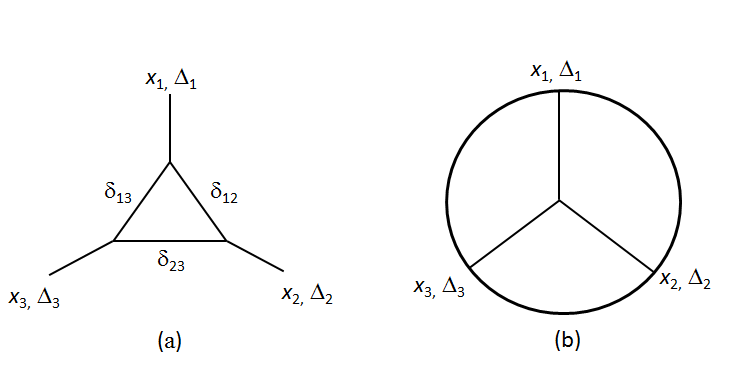}
	\caption{A position space loop integral (a) and its AdS counterpart (b).}
	\label{PosLoop}
\end{figure}

Many other different pictures are possible by splitting the tetrahedra in various ways, or joining some of them onto higher dimensional polytopes. Also, by appropriately varying the dimensions of the internal propagators poles can develop which relate this integral onto simpler ones; such as the 12-pt star. Therefore this part of the story also generalizes. Finally, one may well wonder about position-space loop integrals. These match onto ``window''-type diagrams in the original momentum space. A systematic exploration of these diagrams is beyond our reach, but we can consider the particular example given in figure \ref{PosLoop}. 
The position space integral takes the form
	\bea
	I=\int \prod_{i=1}^3 \frac{\ud Q_i}{2\pi^{d/2}} \prod_{i=1}^3\frac{\Gamma(\Delta_i)}{(P_i\cdot Q_i)^{\Delta_i}}\,\frac{\Gamma(\delta_{12})}{(Q_1\cdot Q_2)^{\delta_{12}}}\frac{\Gamma(\delta_{13})}{(Q_1\cdot Q_3)^{\delta_{13}}}\frac{\Gamma(\delta_{23})}{(Q_2\cdot Q_3)^{\delta_{23}}}
	\eea
where conformality is assumed, e.g. $\Delta_1+\Delta_{12}+\Delta_{13}=d$. The trick now is to realize that
	\bea
	\frac{\Gamma(\delta_{12})}{(Q_1\cdot Q_2)^{\delta_{12}}}\frac{\Gamma(\delta_{13})}{(Q_1\cdot Q_3)^{\delta_{13}}}\frac{\Gamma(\delta_{23})}{(Q_2\cdot Q_3)^{\delta_{23}}}=\iads \ud X\, \prod_{i=1}^3 \frac{\Gamma(\delta_i)}{(Q_i\cdot X)^{\delta_i}}
	\eea
with $\delta_{12}=\frac 12(\delta_1+\delta_2-\delta_3)$, and cyclic permutations thereof. With the three $Q_i$ variables so decoupled, we can use our usual tricks to turn the corresponding integrations into $AdS$ ones. In this case the result is that the triangle gets ``unglued'' onto a star interaction. Since we only included one external point per vertex what we are seeing is three $AdS$ lines meeting at a common, integrated point. Had we included more external points, we would have obtained simplices joined together by one common vertex. This result suggests that perhaps other position-space loop integrals can be similarly unglued onto $AdS$ interactions. However, the story will have to be necessarily more complicated since already at the 4-pt level there is no obvious $AdS$ amplitude which yields the right product of four propagators. 

\section{Discussion}

In this note we have introduced spline technology to quantum field theory. More precisely, we have shown that {\em exact} calculations can be rewritten in terms of splines. The spline seems to act as a bridge between the integrand and its geometrical interpretation, translating identities on one side to the other and vice-versa. We believe that the spline representation will be useful in establishing the dictionnary between integrand and geometry originated in the works of \cite{Hodges:2009hk,ArkaniHamed:2010gg}.

This dictionnary works not only for the fashionable dual-conformal loop integrals of $N=4$ SYM, but in fact for any planar loop-type integral, with varying weights, masses or no masses, numerators, and so on. All these become associated to a certain colored polytope in $AdS$. Beyond one-loop level an intriguing structure appears, whose meaning has so far eluded us - one obtains polytopes with support on other polytopes. However we believe that the ``charge'' interpretation of the spline, to be discussed shortly, will play an important role in elucidating this structure. 

With these results we have seen that a very large class of flat-space integrals can be completely translated into a dual computation in $AdS$ space. But even correlation functions computed directly in $AdS$ using Witten diagrams also fall under the charms of the spline, and we get a beautiful correspondence between star integrals in $AdS$ and their glued up versions as hyperbolic polyhedra. We should note it is not the first time that a connection between Feynman integrals and hyperbolic polytopes has been made. The works \cite{Schnetz:2010pd} and \cite{Davydychev:1997wa} discuss one-loop integrals. Their approach can be summarized and related to ours in the following way. If we begin with the one-loop expression we can write
	\bea
	I&=&\iads \ud X\, \mathcal T(X;\{P_i\})=\int \ud^D X\, \delta(X^2+1) \mathcal T(X;\{P_i\}) \nonumber \\
	&=&\int_0^{+\infty} \prod_{i=1}^n \ud t_i\, \delta( \sum t_i P_{ij} t_j+1).
	\eea
From this point of view it is clear the value of the integral is the same as the volume of a certain ellipsoid in Schwinger parameter space. Depending on the eigenvalues of the metric $P_{ij}$ this can be a spherical or hyperbolic ellipsoid. We can play the same trick with multiloop integrals using our spline expressions but the resulting expressions then no longer have a simple interpretation. 

Besides the theoretical and conceptual questions posed by the existence of the spline representation, we have also endeavoured to show that there are very practical and useful consequences of this representation. One of these regards simple algebraic decompositions of higher point integrals in lower dimensional kinematics, a theme which is currently under investigation \cite{workinprogress}. Another, is the simple derivation of identities between conformal integrals with differing conformal weights. Indeed, augmenting the conformal weight by unity introduces a factor $W\cdot X$ in the spline. Since the splines are being integrated with a gaussian measure, we can trade this factor for a derivative, thereby establishing relations  between integrals with differing conformal weights. For instance, there are many different relations between conformal blocks \cite{Dolan:2000ut,Dolan:2003hv,Dolan:2011dv} with different weights, spin and so on. It is our belief that some of these relations can be rederived from the spline picture. But we do not have to stop there; any one-loop integral will satisfy similar relations which can be explicitly derived from the spline picture. For instance, it is straightforward to check how this works for the $\ _2 F_1$ function, which is related to the 2-node spline, as shown in appendix B, and one indeed recovers the usual transformation properties of the $\ _2F_1$ upon integral shifts. It might be interesting to work out some of these rules for more complicated higher point integrals.

Splines have an important application that, though thus far has been unmentioned, suggests some ideas on its meaning. This application refers to the counting of integer points in polytopes. For physicists, this is an important problem which we can reformulate as follows: in how many ways can a given charge $Q$ be decomposed in terms of a set of elementary charges $q_i$? That is, can we compute the {\em degeneracy}, defined by
	\bea
	d_{\{q_i\}}(Q)=\#\left\{\{n_i\} | \sum_i n_i q_i=Q\right \}.
	\eea
Notice that the charges do not have to be scalars. It is clear that the above is simply a discrete version of the spline, and it counts the number of integer points inside the polytope defined by $M\cdot T=Q$, with $M=(q_1,\ldots q_n)$ the matrix whose columns are the charge vectors. For large enough $Q$, the number of such points is well approximated by the volume of the polytope itself, and the degeneracy becomes precisely the spline!
	\bea
	d_{\{q_i\}}(Q)\rightarrow \mathcal T(Q,\{q_i\})=\int_0^{+\infty} \prod_{i=1}^n \ud t_i \, \delta(Q-\sum t_i q_i)
	\eea
In other words, the spline is the continuous limit of the degeneracy, and in fact it is its leading approximation for large $Q$. Now, how would one go about computing the degeneracy? Physicists have also understood how to do this: one computes a partition function introducing a chemical potential $\mu$,
	\bea
	I(\mu)=\sum_Q d_{\{q_i\}}(Q)e^{-Q\cdot \mu}=\prod_{i=1}^n \frac{1}{1-e^{-q_i\cdot \mu}}
	\eea
and reads off the degeneracy by doing a power expansion. But now we see that in the continuous limit, forming the partition function corresponds simply to computing the Laplace transform, and we know that the transform of the spline is the integrand! And indeed, if we expand the partition function above for small $q_i$ we obtain a product of ``propagators'':
	\bea
	Z(\mu)\rightarrow \prod_{i=1}^n \frac{1}{q_i\cdot \mu}.
	\eea
With this information, what meaning can be given to the integral? We have seen that a one-loop integral can be written as
	\bea
	I=\int_{\mathds M^{d+1,1}} \ud X e^{X^2} \mathcal T(X;\{P_i\}).
	\eea
In our discrete analogy this would be
	\bea
	I=\sum_Q d_{\{q_i\}}(Q)\,e^{-Q^2}.
	\eea
This has again the flavour of a partition function. This is clear if we think of the elementary charges $q_i$ as momenta, since then we can think of $Q^2$ as energy and we have
	\bea
	I=\mbox{Tr}\, e^{-H}
	\eea
This is not a number, but rather a function of the elementary charges, which we can think of as implementing restrictions on the trace.
These results are suggestive that there is a deeper structure behind the spline representation of the loop integrals. For now we leave further investigations for the future.

To end on another speculative note, let us mention the curious similarities between the splines and the twistor space delta functions which are the bread and butter of $N=4$ SYM amplitudes calculations \cite{Adamo:2011pv}. Perhaps the latter could be understood as a complex variable generalization of splines. Then the geometrical interpretation of scattering amplitudes in $N=4$ SYM would be one step closer to being a reality.

\acknowledgments{This work has benefitted from discussions with several people, in particular S. El-Showk, D. Nandan, D. Skinner, M. Spradlin and A. Volovich. We also gratefully acknowledge funding from the LPHTE, Paris, where large part of this work was done. The author is currently supported by D.O.E. - Grant DE-FG02-91ER40688.}

\appendix

\section{Relationship to Carlson's Dirichlet B-splines}
From the Dirichlet spline or generalized truncated power,
	\bea
	\mathcal T_{\{\Delta_i\}}(X,\{P_i\})=\int_0^{+\infty} \prod\frac{\ud t_i}{t_i}\, t_i^{\Delta_i}\, \delta(X-\sum t_i P_i)
	\eea
we can go to Carlson's Dirichlet B-spline \cite{Carlson}, by removing the overall scale. More precisely the $\mathcal T$-spline is supported on a polyhedral cone, so by ``gauge fixing'' the scale, we will get an object supported on a polyhedron. This can be done by introducing a partition of unity
	\bea
	\int \ud v \delta(v-\sum t_i)
	\eea
followed by a rescaling of the $t_i$, $t_i\to v t_i$. The one loop integral becomes
	\bea
	I=\int \ud X\, e^{X^2}\mathcal T_{\{\Delta_i\}}(X;\{P_i\})=\int \frac{\ud X}{X^{\Sigma}} B_{\{\Delta_i\}}(X;\{P_i\})
	\eea
with $\Sigma\equiv \sum_i \Delta_i$ and
	\bea
	B_{\{\Delta_i\}}(X;\{P_i\})=\int \prod_{i=1}^n \frac{\ud t_i}{t_i}\, t_i^{\Delta_i} \delta(1-\sum t_i)\, \delta(X-\sum t_i P_i)
	\eea
Intuitively, going from $\mathcal T$ to $B$ corresponds to taking a flat slice of the polyhedral cone. Indeed, $B$ has support in the convex hull of the vertices $P_i$, or in other words, the polyhedron with vertices $P_i$. Carlson introduces the concept of {\em Dirichlet average} $F$ of a function $f$, which can be formulated as the convolution of a function $f(X)$ with the spline. More precisely, we have: 
	\bea
F(\Delta_i,P_i)=\frac{\Gamma(\Sigma)}{\prod_{i=1}^n \Gamma(\Delta_i)}	\int \ud^{D}X\, f(X)\, B_{\{\Delta\}}(X,\{P_i\})
	\eea
The one-loop integral is then the Dirichlet average of a power function, which is one of the better studied examples. Carlson defines the resulting function as an $R$-hypergeometric function. Some elementary properties of the $B$-spline which pass straightforwardly onto the $\mathcal T$-spline is that when knots coalesce, they can be replaced by a single one, with a weight corresponding to the sum of the weights of the previous ones, e.g.
	\bea
	F(\underbrace{\{\Delta_1,\Delta_2,\ldots, \Delta_n\}}_n,\underbrace{\{P_1,P_1,\ldots,P_n\}}_n)=F(\underbrace{\{\Delta_1+\Delta_2,\ldots, \Delta_n\}}_{n-1},\underbrace{\{P_1,\ldots,P_n\}}_{n-1})
	\eea
Also, by analytic continuation, the requirement that the $\Delta_i$ have positive real part can be relaxed. In this case a knot with zero weight may be omitted.

\section{A two-loop computation}

We start off with the double box integral, or exchange diagram in position space
	\bea
	I_{n,m}=\int \frac{\ud^d Q_1 \ud^d Q_2}{(2\pi^{d/2})^2}\, \frac{\Gamma(\delta)}{(-\,Q_1\cdot Q_2)^\delta}\, \prod_{i\in L} \frac{\Gamma(\Delta_i)}{(- P_i\cdot Q_1)^{\Delta_i}}\prod_{j\in R} \frac{\Gamma(\Delta_j)}{(- P_j\cdot Q_2)^{\Delta_j}}
	\eea
\begin{figure}
	\centering
		\includegraphics[width=10 cm]{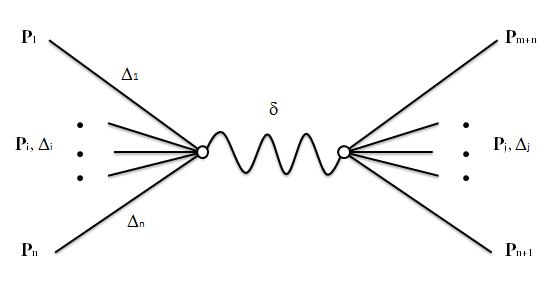}
	\label{fig:exchangediagram}
\end{figure}
Introducing Schwinger parameters and splines we can write
	\bea
	I_{n,m}=\int_{M^D} \ud X_1  \ud X_2\, \mathcal T_{S_L}(X_1)\mathcal T_{S_R}(\bar X_2)\, \int \frac{\ud^d Q_1 \ud^d Q_2}{(2\pi^{d/2})^2} \int_0^{+\infty} \frac{\ud s}s\, s^{\delta} e^{ Q_1\cdot X_1+ Q_1\cdot X_2+ s\, Q_1 \cdot Q_2}
	\eea
where $\mathcal T_{S}(X)$ denote the splines for the simplices $S$, which are described not only by their vertices $P_i$ but also by a coloring of these vertices given by the weights $\Delta_i$.  To proceed we first notice that the overall scales $|X_1|$ and $|X_2|$ can be decoupled from the splines; since the splines have support on the cone formed by the null vectors $P_i$, the integrals have support only where $X_i^2<0$, so this is well defined. Indeed we have 
	\bea
	\mathcal T_{S_L}(\bar X_1)=| X_1|^{\Sigma_L-D}\, \mathcal T_{S_L}(X_1)=| X_1|^{-\delta-2}\, \mathcal T_{S_L}(X_1)
	\eea
with $X_1^2=-1$, and where we used the conformality condition $\Sigma_L+\delta=d=D-2$. Defining the integral over the coloured hyperbolic simplex by 	
	\be
	\int_{S}\ud X\equiv \int_{\mathds M^D} \ud  X \, \delta(X^2+1)\, \mathcal T_{S}(X)
	\ee
Then we can write
	\bea
	I_{n,m}=\int_{S_L}\ud X_1\int_{S_R} \ud X_2 \, \bar G_\delta(X_1,X_2)
	\eea
with the convolving kernel
	\bea
	\bar G_\delta(X_1,X_2)=
	\int_0^{+\infty} \frac{\ud v_1}{v_1} \frac{\ud v_2}{\ud v_2}\frac{\ud s}{s} (v_1 v_2)^{d-\delta}\, s^\delta  \int \frac{\ud Q_1 \ud Q_2}{(2\pi^{d/2})^2}  e^{2 v_1 Q_1\cdot X_1+2 v_2 Q_1\cdot X_2+2 s\, Q_1 \cdot Q_2}
	\eea
Notice that at this point we have reduced the computation of $I_{n,m}$ to a double integral over $AdS$ space. It is our next job to figure out what this convolving kernel is. Performing first the $Q_1$ and $Q_2$ integrals we find
	\bea
	\bar G_\delta(X_1,X_2)=\,\int_0^{+\infty} \frac{\ud v_1}{v_1} \frac{\ud v_2}{\ud v_2}\frac{\ud s}{s} (v_1 v_2)^{d-\delta}\, s^\delta\, e^{-(1+s^2)v_1^2-\,v_2^2+2\,s\, v_1 v_2\, X_1\cdot X_2}
	\eea
Doing rescalings $s\to s/v_1$ the $v_1$ integral is simply performed. Renaming $v_2\to \bar s$ and defining the chordal distance $u\equiv (X_1-X_2)^2$ we find
	\bea
\bar G_\delta(X_1,X_2)=\frac{\Gamma\left(\frac{d-2\delta}2\right)}2\int_0^{+\infty} \frac{\ud s}{s}\, \frac{\ud \bar s}{\bar s}\,s^{\delta}s^{d-\delta}\, e^{-(s+\bar s)^2-s \bar s u}
	\eea
This is remarkably close to the expression for the bulk-to-bulk propagator of a scalar of dimension $\delta$ in $AdS_{d+1}$. Notice that 
this expression follows from
	\bea
	\bar G_\delta(X_1,X_2)=\gt{d-2\delta}\int \frac{\ud Q}{2\pi^{d/2}}\, \frac{\Gamma(\delta)}{(X_1\cdot Q)^{\delta}}\frac{\Gamma(d-\delta)}{(X_2\cdot Q)^{d-\delta}}
	\eea
Continuing,
	\bea
	\bar G_\delta(u)&=& 2 \,\Gamma\left(\frac{d-2\delta}2\right)\oint \frac{\ud c}{2\pi i}\, \Gamma(c) u^{-c}\, 
\int_0^{+\infty} \frac{\ud s}{s}\, \frac{\ud \bar s}{\bar s}\,s^{\delta-c}s^{d-\delta-c}\,  e^{-(s+\bar s)^2-s \bar s u}= \nonumber \\
&=& \Gamma\left(\frac{d-2\delta}2\right)\oint \frac{\ud c}{2\pi i}\, \frac{\Gamma(c) \Gamma(d/2-c)\Gamma(\delta-c)\Gamma(d-\delta-c)}{\Gamma(d-2c)}\, u^{-c} \nonumber\\
&=& \frac{\sqrt{\pi}}{2^{d-1}}\, \Gamma\left(\frac{d-2\delta}2\right)\oint \frac{\ud c}{2\pi i}\,\frac{\Gamma(c)\Gamma(\delta-c)\Gamma(d-\delta-c)}{\Gamma(\frac 12+\frac d2-c)}\, \left(\frac u4\right)^{-c}\nonumber \\
&=& \frac{\sqrt{\pi}}{2^{d-1}}\, \frac{\gt{d-2\delta}\gn{\delta}\gn{d-\delta}}{\gt{1+d}}\, _2 F_1\left(\delta,d-\delta,\frac 12+\frac d2, -\frac{u}4\right)
\eea
Now we can use a hypergeometric identity:
	\bea
	 _2  F_1(a,b,c,z)&=&\left(\frac{\gn{b-a}\gn{c}}{\gn{b}\gn{c-a}}\right) (-z)^{-a}\ _2 F_1(a,1+a-c,1+a-b,\frac 1z)\nonumber \\
	&+& \left(\frac{\gn{a-b}\gn{c}}{\gn{a}\gn{c-b}}\right) (-z)^{-b}\ _2 F_1(b,1+b-c,1+b-a,\frac 1z)
	\eea
which holds whenever $b\neq a+n$ for all $n\in \mathds Z$, and $z\notin (0,1)$. To us this imposes that $\delta\neq d/2+n$. When this is true, we can use this identity to find
	\bea
	\bar G_{\delta(u)}&=&\frac{\sqrt{\pi}}{2^{d-1}} \,\gt{d-2\delta}\,\left\{ \left(\frac{\gn{d-2\delta} \Gamma(\delta)}{\gt{1+d-2\delta}}\right)\, G_{\delta}(u)+
		\left(\frac{\gn{2\delta-d} \Gamma(d-\delta)}{\gt{1+2\delta-d}}\right)\, G_{d-\delta}(u)\right\} \nonumber \\
	&=& \frac{\sqrt{\pi}}{2^{d-1}} \,\gt{d-2\delta}\, \left\{\, 2^{d-2\delta-1} \gt{d-2\delta}\gn{\delta}\,G_{\delta}(u)+2^{2\delta-d-1}\, \gt{2\delta-d}\gn{d-\delta}\, G_{d-\delta}(u)\right\} \nonumber
	    \eea
where
\bea
G_{\Delta}(u)\equiv \frac{1}{u^\Delta}\, _2 F_1\left (\Delta,\frac{1+2\Delta-d}2,1+2\Delta-d,-\frac 4u\right)
\eea
is, up to a normalization factor, the bulk-to-bulk propagator for a field of dimension $\Delta$. This shows that the double box integral has a nice geometric interpretation: it corresponds to an integral over two simplices glued by the exchange of scalars propagating in $AdS$.

\end{document}